\documentclass[%
 reprint,twocolumn,
 amsmath,amssymb,
 aps,
]{revtex4-2}

\usepackage{graphicx}
\usepackage{dcolumn}
\usepackage{bm}
\usepackage{braket}
\usepackage{physics}
\usepackage{siunitx}

\begin{document}

\preprint{APS/123-QED}

\title{Coupled Three-Mode Squeezed Vacuum}

\author{Wenlei Zhang}
\author{Ryan T. Glasser}%
 \email{rglasser@tulane.edu}
\affiliation{%
 Department of Physics and Engineering Physics, Tulane University, New Orleans, Louisiana 70118, USA
}%




\date{\today}

\begin{abstract}

Multipartite entanglement is a key resource for various quantum information tasks. Here, we present a scheme for generating genuine tripartite entanglement via nonlinear optical processes. We derive, in the Fock basis, the corresponding output state which we termed the \textit{coupled three-mode squeezed vacuum}. We find unintuitive behaviors arise in intensity squeezing between two of the three output modes due to the coupling present. We also show that this state can be genuinely tripartite entangled.

\end{abstract}

\maketitle

Quantum entanglement has been a topic of extensive research ever since Einstein, Podolsky, and Rosen proposed the nonlocal nature of quantum mechanics \cite{einstein_can_1935}. More recently, entanglement theory has been generalized to systems with more than two subsystems \cite{greenberger_going_2007,dur_three_2000,briegel_persistent_2001}. Such multipartite entanglement, integral to the generation of future quantum networks, has also been observed experimentally \cite{bouwmeester_observation_1999,pan_experimental_2000}.

Quantum networks enable quantum communication between arbitrary network users \cite{wehner_quantum_2018}. Multipartite entanglement serves as a fundamental resource for various quantum information tasks in such networks \cite{mccutcheon_experimental_2016}. For example, information exchange in a network may be done through quantum teleportation \cite{zeilinger_quantum_2018}, which requires prior generation and distribution of entanglement between multiple parties. Multipartite entanglement also allows parties in the network to perform distributed information-processing tasks that exhibit quantum advantages \cite{buhrman_nonlocality_2010}. One of the many obstacles in realizing quantum networks is precisely the generation and distribution of robust multipartite entanglement.

Nonlinear optics provides a number of promising experimental tools for realizing multipartite entanglement, one of which is four-wave mixing (FWM).  Four-wave mixing is a general term used to describe the parametric interaction between four coherent fields mediated by a nonlinear medium. Theoretical studies on using FWM for the generation of a two-mode squeezed vacuum (TMSV) or other two-mode squeezed states have been well-presented \cite{yuen_generation_1979,bondurant_pump_1984,kumar_squeezed-state_1984,reid_generation_1985,bondurant_degenerate_1984,gerry_introductory_2005}. FWM in hot atomic vapor has been demonstrated to be a reliable source of squeezed light experimentally \cite{mccormick_strong_2007,guo_experimental_2014,gupta2016,swaim_squeezed-twin-beam_2017}. Recent work has utilized FWM as an efficient source of multimode quantum correlated states using either multiple pump beams \cite{jia_generation_2017,Cai_Hao_Zhang_Liu_Luo_Zheng_Li_Zhang_2020,Dong_Zhang_Jing_2022}, spatially structured pump beams \cite{wang_single-step_2017,swaim_multimode_2018}, or cascading setups \cite{wang_generation_2016,wang_characterization_2017,cao_experimental_2017,cai_quantum-network_2015}. Here, we show that such processes (as well as those like spontaneous parametric down conversion) may be used to generate genuine tripartite entangled states and derive the generated quantum state.

\begin{figure}[htbp]
    \centering
    \includegraphics[width=\linewidth]{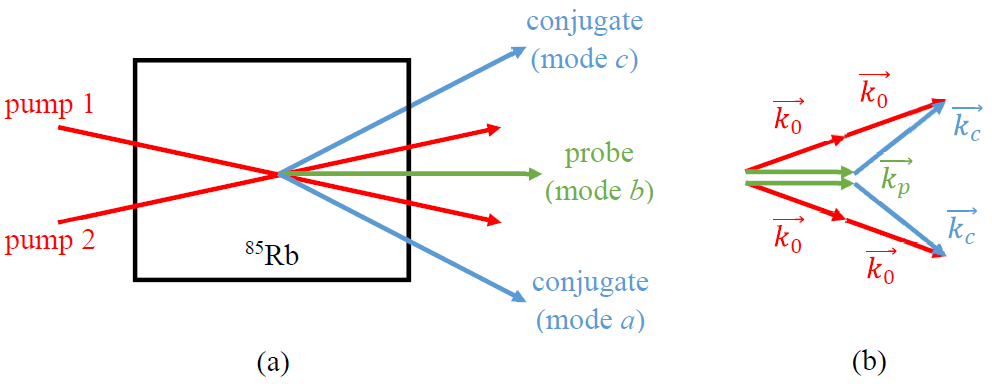}
    \caption{\label{fig:dualpump}Dual-pump coupled FWM. (a) Schematic of setup. (b) Phase-matching diagram. $\protect\overrightarrow{k_0}$, $\protect\overrightarrow{k_p}$, $\protect\overrightarrow{k_c}$ are the wavevectors of the pump, probe, and conjugate photons, respectively.}
\end{figure}

In this letter, we consider the setup in Fig.~\ref{fig:dualpump} (a). The pump photons interact with the nonlinear medium (\textsuperscript{85}Rb atomic vapor) through a FWM process. The two pump beams of the same frequency are aligned such that their output probe beams overlap in the middle (mode $b$). In this configuration, two photons from pump 1 are annihilated to create a probe photon in mode $b$ and a conjugate photon in mode $a$, and two photons from pump 2 are annihilated to create a probe photon in mode $b$ and a conjugate photon in mode $c$. Figure \ref{fig:dualpump} (b) shows the phase-matching diagram.

The time-independent interaction Hamiltonian for this setup can be written as
\begin{equation}
    \hat{H}_I=i\hbar(\eta_1^\ast\hat{a}\hat{b}-\eta_1\hat{a}^\dagger\hat{b}^\dagger+\eta_2^\ast\hat{b}\hat{c}-\eta_2\hat{b}^\dagger\hat{c}^\dagger),
\end{equation}
where $\hat{a}$, $\hat{b}$, $\hat{c}$ are the bosonic annihilation operators for modes $a$, $b$, $c$, respectively, $\eta_1$ and $\eta_2$ are complex numbers that depend on the respective states of pumps 1 and 2 and the third-order nonlinearity of the medium. Here we assume undepleted and coherent pumps, and make use of the rotating wave approximation \cite{scully_quantum_1997}. The unitary time evolution operator associated with this Hamiltonian is what we call the \textit{coupled three-mode squeezing operator}:
\begin{gather}
    \hat{S}_{3}(\xi_{1},\xi_{2})=\operatorname{exp}(\xi_{1}^\ast\hat{a}\hat{b}+\xi_{2}^\ast\hat{b}\hat{c}-\xi_{1}\hat{a}^{{\dagger}}\hat{b}^{{\dagger}}-\xi_{2}\hat{b}^{{\dagger}}\hat{c}^{{\dagger}}), \\
    \xi_1=\eta_1t=r_1e^{i\theta_1},\;\xi_2=\eta_2t=r_2e^{i\theta_2},
\end{gather}
where $t$ is the interaction time. This interaction can be characterized by the two complex squeezing parameters $\xi_1$ and $\xi_2$, whose respective magnitudes are $r_1$, $r_2$, and respective phases are $\theta_1$, $\theta_2$. We note that this squeezing operator can also be realized using other nonlinear optical processes such as spontaneous parametric down-conversion.

The \textit{coupled three-mode squeezed vacuum} (C3MSV) is the output state when $\hat{S}_3$ acts on vacuum inputs. The result is, in the Fock basis, 
\begin{widetext}
\begin{equation}
    \ket{\text{C3MSV}}\equiv\hat{S}_{3}\ket{0,0,0}=\frac{1}{\cosh{r}}\sum_{n,l=0}^{\infty}(-1)^{n+l}e^{i(n\theta_{1}+l\theta_{2})}\Big(\frac{r_{1}}{r}\tanh{r}\Big)^{n}\Big(\frac{r_{2}}{r}\tanh{r}\Big)^{l}\sqrt{\frac{(n+l)!}{n!\,l!}}\ket{n,n+l,l},
\end{equation}
\end{widetext}
where $r=\sqrt{r_1^2+r_2^2}$, and the set of states $\{\ket{n,m,l}\equiv\ket{n}_{a}\otimes\ket{m}_{b}\otimes\ket{l}_{c}\}$ is the basis of the three-mode Fock space. The C3MSV becomes a TMSV in modes $a$, $b$ and vacuum in mode $c$ when $r_1>0$ and $r_2=0$, and a TMSV in modes $b$, $c$ and vaccum in mode $a$ when $r_1=0$ and $r_2>0$.

\begin{figure}[htbp]
    \centering
    \includegraphics[width=\linewidth]{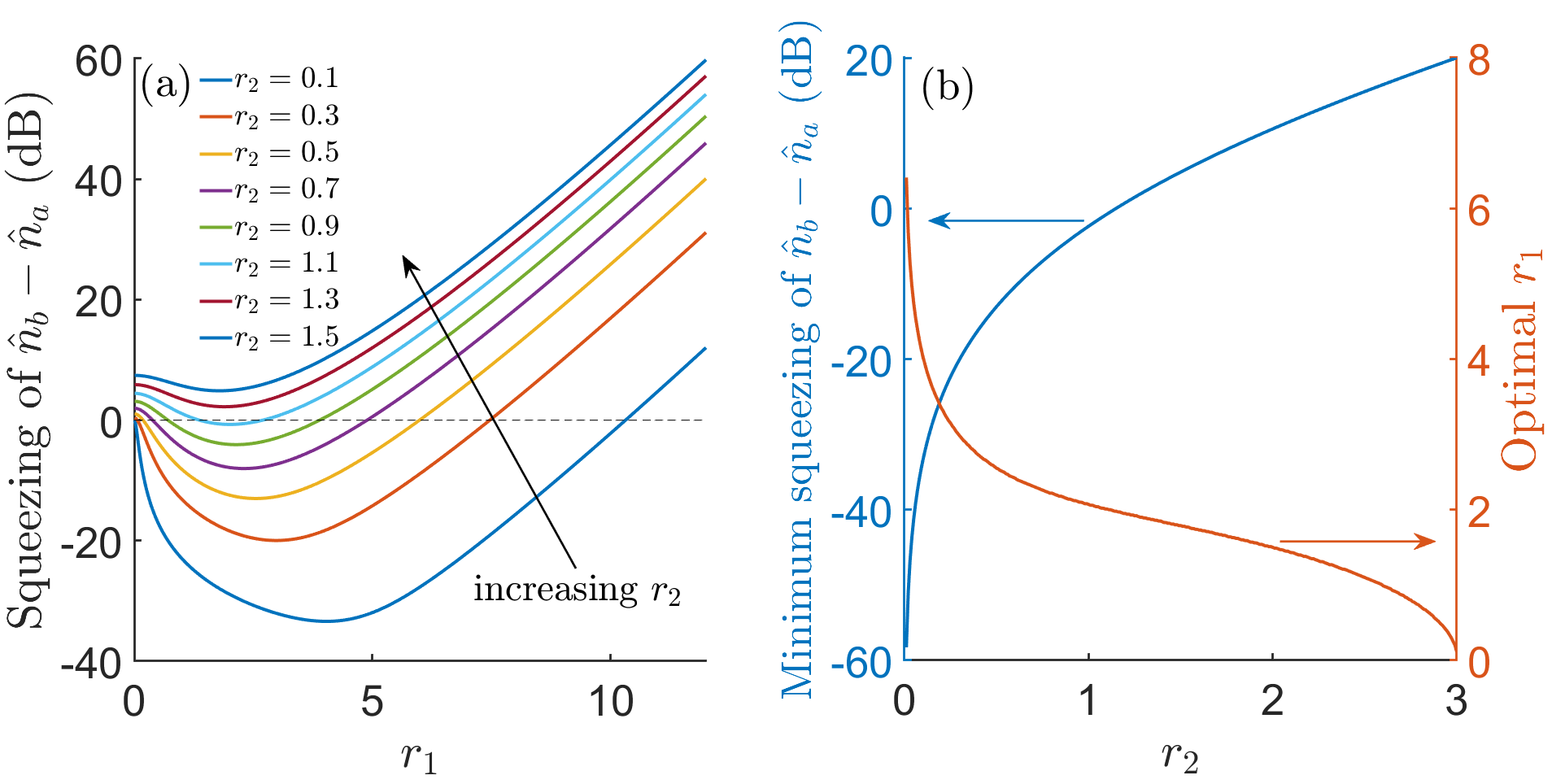}
    \caption{\label{fig:intsquz}(a) Squeezing, in decibels, of the photon number operator $\hat{n}_b-\hat{n}_a$ with changing $r_1$ and $r_2$. (b) Minimum value of squeezing of $\hat{n}_b-\hat{n}_a$ and the value of $r_1$ where the minimum occurs w.r.t $r_2$.}
\end{figure}

Having found the output state, we now look at the intensity squeezing properties of said state, namely the squeezing of combinations of photon number operators $\hat{n}_a\equiv\hat{a}^{\dagger}\hat{a}$, $\hat{n}_b\equiv\hat{b}^{\dagger}\hat{b}$, and $\hat{n}_c\equiv\hat{c}^{\dagger}\hat{c}$. Squeezing of a Hermitian operator $\hat{O}$ is defined as the ratio $\langle(\Delta\hat{O})^2\rangle/\langle(\Delta\hat{O})^2\rangle_{\text{coh}}$, expressed in decibels, where $\langle(\Delta\hat{O})^2\rangle$ denotes the variance of $\hat{O}$ w.r.t the C3MSV, and $\langle(\Delta\hat{O})^2\rangle_{\text{coh}}$ denotes the standard or shot noise limit (SNL) of $\hat{O}$, which is calculated to be the variance of the operator w.r.t coherent states in each mode with the same respective mean photon number as the C3MSV. A physical quantity is squeezed whenever the squeezing of the corresponding operator is below \SI{0}{dB}.

The C3MSV is an eigenstate of the photon number operator $\hat{n}_b-\hat{n}_a-\hat{n}_c$ with eigenvalue 0. Hence, the variance of the difference between the intensity of mode $b$ and the sum of intensities of mode $a$ and $c$ is ideally 0. This is expected by simple examination of the setup. Interesting behaviors arise when we look at the intensity squeezing between two modes, namely $\hat{n}_b-\hat{n}_a$ and $\hat{n}_b-\hat{n}_c$. Since $r_1$ is the squeezing parameter that characterizes the interaction between pump 1 and the medium, which generates photons in modes $a$ and $b$, one would expect the strength of the squeezing of $\hat{n}_b-\hat{n}_a$ to increase (i.e. the ratio $\langle[\Delta(\hat{n}_b-\hat{n}_a)]^2\rangle/\langle[\Delta(\hat{n}_b-\hat{n}_a)]^2\rangle_\text{coh}$ decreases) as $r_1$ increases, as it would in a TMSV \cite{gerry_introductory_2005}. However, Fig.~\ref{fig:intsquz} (a) shows that a maximum exists for the strength of squeezing of $\hat{n}_b-\hat{n}_a$ (minimum of the ratio $\langle[\Delta(\hat{n}_b-\hat{n}_a)]^2\rangle/\langle[\Delta(\hat{n}_b-\hat{n}_a)]^2\rangle_\text{coh}$) for any value of $r_2\neq0$, which means that the intensity difference between modes $b$ and $a$ will always become not squeezed for sufficiently large $r_1$. Fig.~\ref{fig:intsquz} (b) shows how the minimum of $\langle[\Delta(\hat{n}_b-\hat{n}_a)]^2\rangle/\langle[\Delta(\hat{n}_b-\hat{n}_a)]^2\rangle_\text{coh}$ and the optimal $r_1$ (the value of $r_1$ at the minimum) changes with $r_2$. Each mode in the C3MSV is in a thermal state individually, which is similar to those of a TMSV \cite{schumaker_new_1985,gerry_introductory_2005}. Thus the interaction between modes $a$ and $b$ is effectively being seeded by a thermal state in mode $b$, which is being generated from the interaction between modes $b$ and $c$.

\begin{figure}[htbp]
    \centering
    \includegraphics[width=\linewidth]{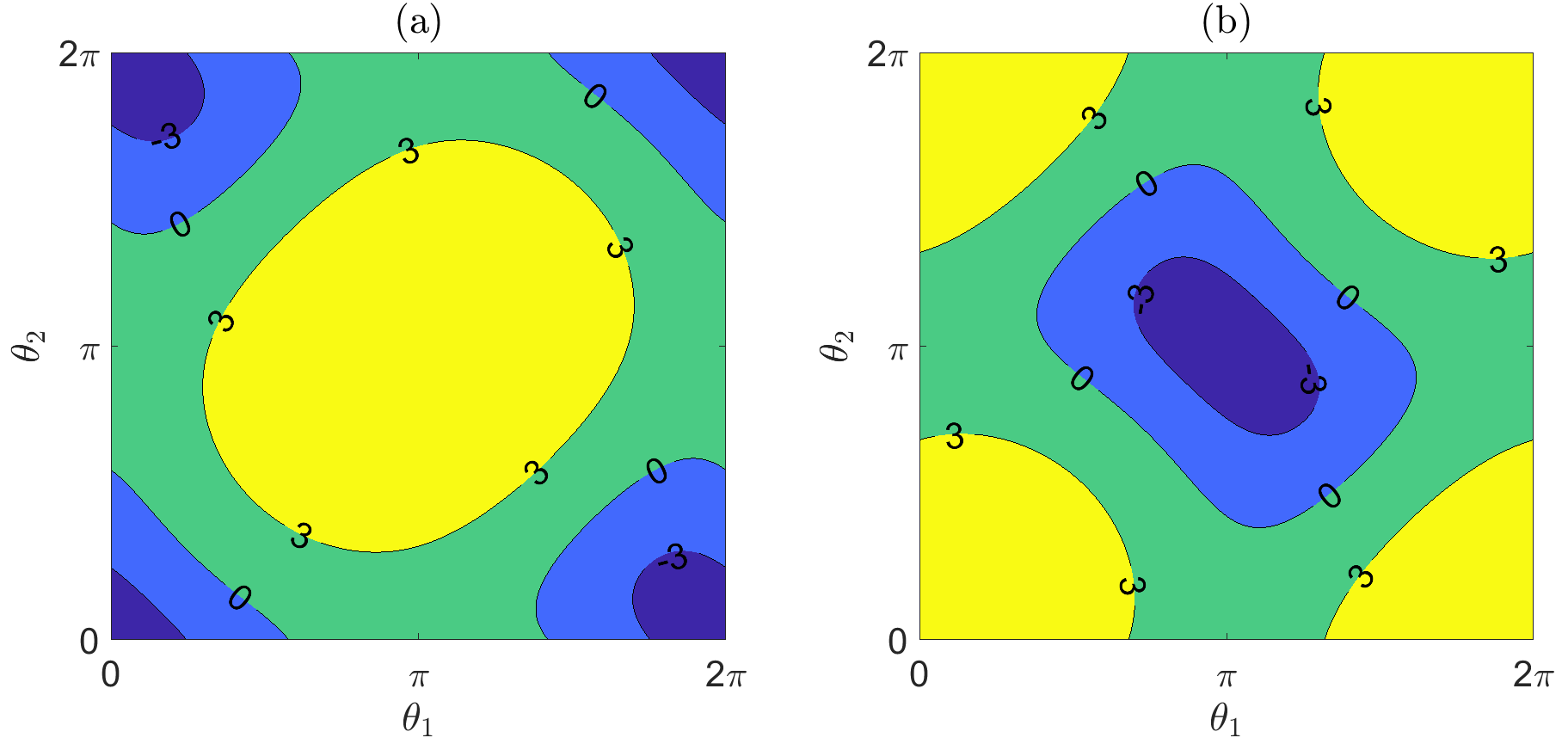}
    \caption{(Color online) Contour plot of squeezing, in decibels, of (a) $\hat{U}$, (b) $\hat{V}$, for $h_1=h_2=h_3=1$, $r_1=r_2=0.5$. The blue regions are where the operators are squeezed.}
    \label{fig:quadrature1}
\end{figure}

\begin{figure}[htbp]
    \centering
    \includegraphics[width=\linewidth]{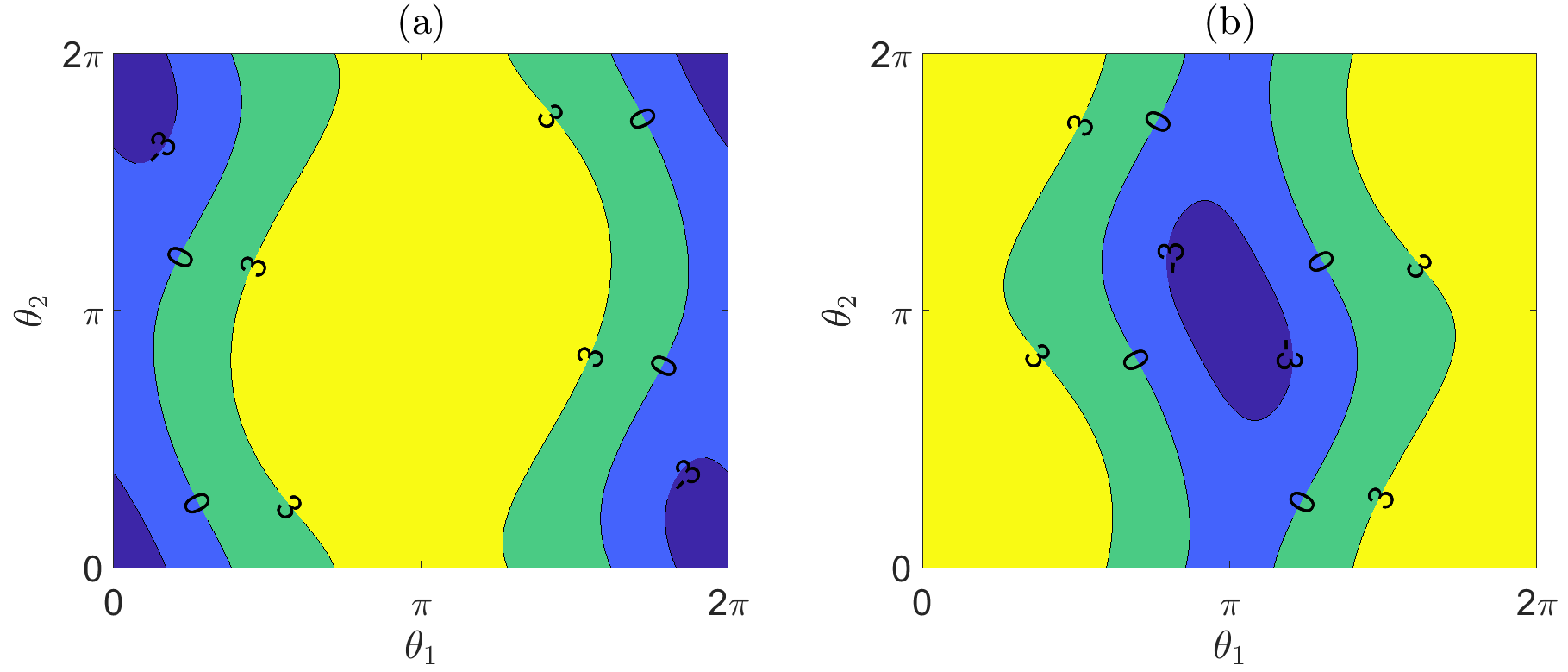}
    \caption{(Color online) Contour plot of squeezing, in decibels, of (a) $\hat{U}$, (b) $\hat{V}$, for $h_1=h_2=h_3=1$, $r_1=0.7$, $r_2=0.3$. The blue regions are where the operators are squeezed.}
    \label{fig:quadrature2}
\end{figure}

The intensity squeezing results shown above only depend on the magnitudes $r_1$ and $r_2$, but not on the phases $\theta_1$ and $\theta_2$. Experimentally, $r_1$ and $r_2$ depend on factors such as temperature of the medium, pump powers, and interaction time, so their tunability is limited. To investigate the presence of entanglement, we would like to look at the quadrature squeezing properties of the C3MSV, which do depend on the phases $\theta_1$ and $\theta_2$. Here, we introduce the quadrature operators
\begin{gather}
    \hat{U}=h_1\hat{X}_a+h_2\hat{X}_b+h_3\hat{X}_c, \\
    \hat{V}=h_1\hat{P}_a+h_2\hat{P}_b+h_3\hat{P}_c, 
\end{gather}
where $\hat{X}_j=(\hat{j}+\hat{j}^{\dagger})/2$, $\hat{P}_j=(\hat{j}-\hat{j}^{\dagger})/(2i)$ for $j=a,b,c$, and $h_1$, $h_2$, $h_3$ are real numbers. The uncertainty principle gives us
\begin{gather}
    \langle(\Delta\hat{U})^2\rangle\langle(\Delta\hat{V})^2\rangle\geq\frac{1}{16}(h_1^2+h_2^2+h_3^2)^2,
\end{gather}
and the SNLs of $\hat{U}$ and $\hat{V}$ are
\begin{gather}
    \langle(\Delta\hat{U})^2\rangle_{\text{coh}}=\langle(\Delta\hat{V})^2\rangle_{\text{coh}}=\frac{1}{4}(h_1^2+h_2^2+h_3^2).
\end{gather}
Figure \ref{fig:quadrature1} and Fig.~\ref{fig:quadrature2} show the squeezing of $\hat{U}$ and $\hat{V}$ for $h_1=h_2=h_3=1$ and various values of $r_1$ and $r_2$. Figure \ref{fig:quadrature1} shows the results for symmetric magnitudes of the squeezing parameters ($r_1=r_2$), and Fig.~\ref{fig:quadrature2} shows the results for asymmetric magnitudes ($r_1{\neq}r_2$). In both cases, there are values of $\theta_1$ and $\theta_2$ where $\hat{U}$ and $\hat{V}$ are both not squeezed, but never were they both squeezed. This is similar to the result of a joint quadrature measurement on a TMSV. For the asymmetric case shown in Fig.~\ref{fig:quadrature2} ($r_1>r_2$), there are certain values of $\theta_1$ where the squeezing is independent of $\theta_2$.

We would also like to show that the C3MSV is \textit{genuine tripartite entangled}. Genuine tripartite entanglement is a stronger criterion than, and is not equivalent to \textit{full tripartite inseparability} in general, the details of which are explained in Ref.~[\onlinecite{shalm_three-photon_2013}]. A criterion based on quadrature operators for demonstrating genuine tripartite entanglement has been presented and proven in Ref.~[\onlinecite{armstrong_multipartite_2015}]. The criterion is stated as follows: the violation of \textit{any} of the inequalities,
\begin{subequations} \label{eq:criterion}
\begin{gather}
    \langle(\Delta\hat{U}_1)^2\rangle\langle(\Delta\hat{V}_1)^2\rangle\geq1, \label{criterion1} \\
    \langle(\Delta\hat{U}_2)^2\rangle\langle(\Delta\hat{V}_2)^2\rangle\geq1, \label{criterion2} \\
    \langle(\Delta\hat{U}_3)^2\rangle\langle(\Delta\hat{V}_3)^2\rangle\geq1, \label{criterion3}
\end{gather}
\end{subequations}
is \textit{sufficient} to confirm genuine tripartite entanglement, where
\begin{gather}
    \hat{U}_1=2\hat{X}_a-\sqrt{2}(\hat{X}_b+\hat{X}_c),\;\hat{V}_1=2\hat{P}_a+\sqrt{2}(\hat{P}_b+\hat{P}_c), \notag \\
    \hat{U}_2=2\hat{X}_b-\sqrt{2}(\hat{X}_a+\hat{X}_c),\;\hat{V}_2=2\hat{P}_b+\sqrt{2}(\hat{P}_a+\hat{P}_c), \notag \\
    \hat{U}_3=2\hat{X}_c-\sqrt{2}(\hat{X}_a+\hat{X}_b),\;\hat{V}_3=2\hat{P}_c+\sqrt{2}(\hat{P}_a+\hat{P}_b). \notag
\end{gather}
Notice that $[\hat{U}_j,\hat{V}_j]=0$, so the product $\langle(\Delta\hat{U}_j)^2\rangle\langle(\Delta\hat{V}_j)^2\rangle$ is lower bounded by $0$ for ${j}=a,b,c$. We find that the inequalities in Eqs.~(\ref{criterion1}) and (\ref{criterion3}) are never violated, but the inequality in Eq.~(\ref{criterion2}) can be violated.
\begin{figure}[htbp]
    \centering
    \includegraphics[width=\linewidth]{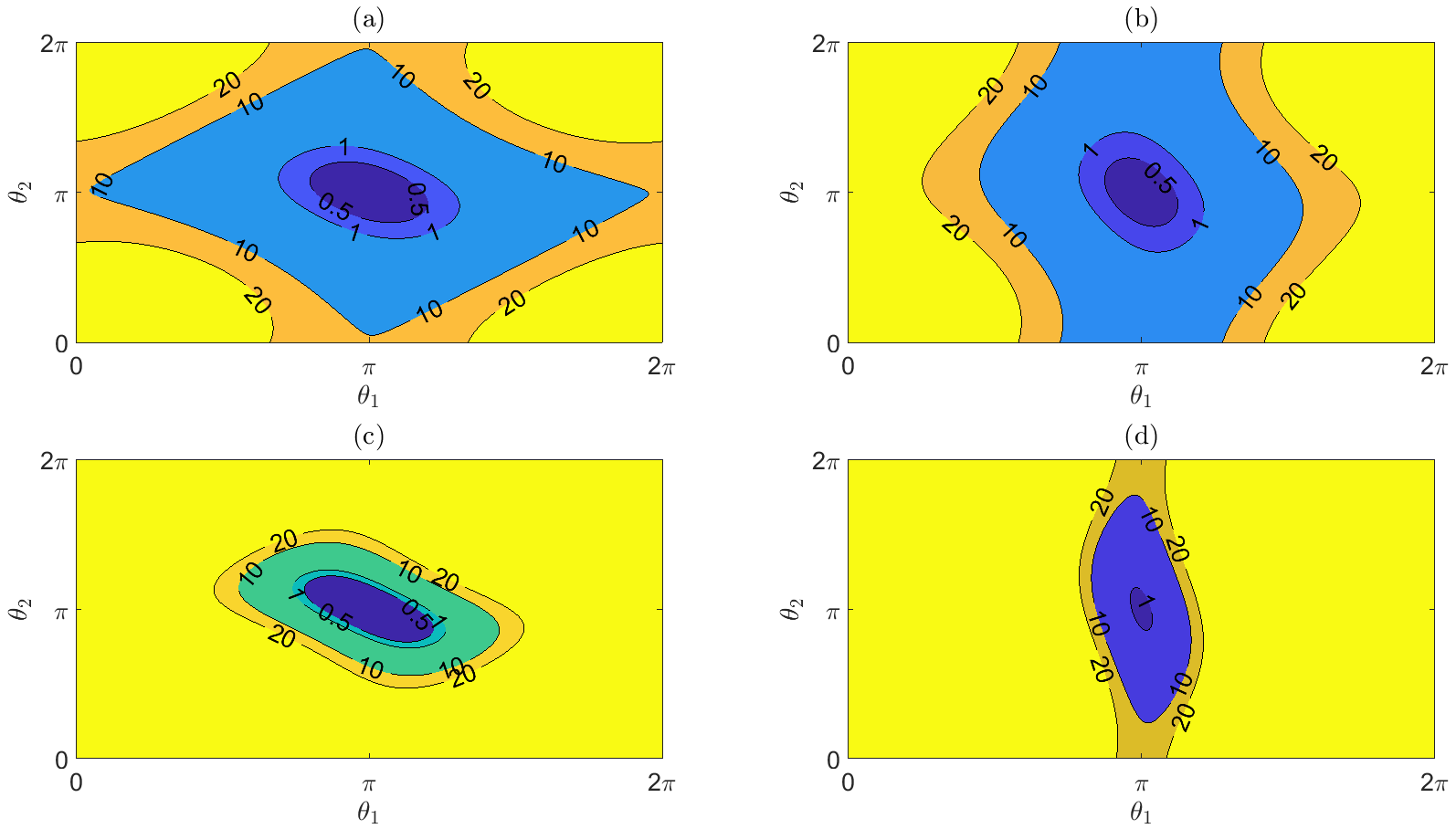}
    \caption{(Color online) Contour plot of the product $\langle(\Delta\hat{U}_2)^2\rangle\langle(\Delta\hat{V}_2)^2\rangle$ for (a) $r_1=r_2=0.5$, (b) $r_1=0.7$, $r_2=0.3$, (c) $r_1=r_2=1.0$, (d) $r_1=1.7$, $r_2=0.3$. The region inside the line labeled ``$1$" is where the inequality in Eq. (\ref{criterion2}) is violated.}
    \label{fig:entanglement}
\end{figure}

Figure \ref{fig:entanglement} shows contour plots of the product $\langle(\Delta\hat{U}_2)^2\rangle\langle(\Delta\hat{V}_2)^2\rangle$ for symmetric and asymmetric values of $r_1$ and $r_2$. In all four cases in Fig.~\ref{fig:entanglement}, the region where we can confirm genuine tripartite entanglement is centered at the point $\theta_1=\theta_2=\pi$. The violation of the inequality in Eq.~\ref{criterion2} becomes stronger at the expense of a smaller range of $\theta_1$ and $\theta_2$ values as $r_1$ and $r_2$ increase. This region becomes smallest in the most asymmetric case (Fig.~\ref{fig:entanglement} (d)) since the C3MSV reduces to a TMSV in modes $a$ and $b$ and vacuum in mode $c$ when $r_1{\gg}r_2$, which is not genuine tripartite entangled.

In conclusion, we presented a scheme for generating tripartite entanglement via nonlinear optical processes, such as FWM, using two crossed pump beams. We derived the output state of this setup for vacuum inputs, which we termed the coupled three-mode squeezed vacuum. We investigated the intensity squeezing and quadrature squeezing properties of the C3MSV state, and showed that some unintuitive behaviors arise due to the coupling. We also showed that C3MSV can be genuinely tripartite entangled. We would like to point out that the violation of inequalities in Eq.~(\ref{eq:criterion}) is only sufficient in demonstrating genuine tripartite entanglement, thus the state may still be genuine tripartite entangled even in regimes where none of the inequalities in Eq.~(\ref{eq:criterion}) are violated. The setup presented here can be modified and expanded to generate more than three output modes \cite{wang_single-step_2017,knutson_optimal_2018}. Such setups provide us with a platform to study multipartite entanglement, and will further the development of quantum information protocols and quantum networks.

\bibliography{c3msv}

\end{document}